\def\KCleanVersion{}
\documentclass[11pt,reqno]{amsart}

\usepackage[T1]{fontenc}
\usepackage{lmodern}
\input{glyphtounicode}
\pdfglyphtounicode{mu}{03BC}
\pdfglyphtounicode{vextendsingle}{23D0}
\pdfglyphtounicode{vextenddouble}{2016}
\pdfglyphtounicode{parenleftbigg}{0028}
\pdfglyphtounicode{parenrightbigg}{0029}
\pdfglyphtounicode{parenleftBigg}{0028}
\pdfglyphtounicode{parenrightBigg}{0029}
\pdfglyphtounicode{braceleftBigg}{007B}
\pdfglyphtounicode{parenlefttp}{239B}
\pdfglyphtounicode{parenleftbt}{239D}
\pdfglyphtounicode{parenrighttp}{239E}
\pdfglyphtounicode{parenrightbt}{23A0}
\pdfglyphtounicode{summationtext}{2211}
\pdfglyphtounicode{summationdisplay}{2211}
\pdfglyphtounicode{producttext}{220F}
\pdfglyphtounicode{productdisplay}{220F}
\pdfglyphtounicode{tildewide}{0303}
\pdfglyphtounicode{radicalbig}{221A}
\pdfglyphtounicode{radicalBig}{221A}

\usepackage{amsmath,amssymb,amsthm}
\usepackage{microtype}
\usepackage{booktabs,tabularx}
\usepackage{xcolor}
\definecolor{revisionred}{RGB}{180,25,25}
\newif\ifshowrevisions
\ifdefined\KCleanVersion\showrevisionsfalse\else\showrevisionstrue\fi
\newcommand{\revision}[1]{{\ifshowrevisions\color{revisionred}\else\color{black}\fi#1}}
\newenvironment{revisionblock}
  {\par\begingroup\ifshowrevisions\color{revisionred}\else\color{black}\fi}
  {\par\endgroup}
\usepackage[hidelinks]{hyperref}

\newtheorem{theorem}{Theorem}[section]
\newtheorem{proposition}[theorem]{Proposition}
\newtheorem{lemma}[theorem]{Lemma}
\newtheorem{corollary}[theorem]{Corollary}
\theoremstyle{definition}
\newtheorem{definition}[theorem]{Definition}
\theoremstyle{remark}
\newtheorem{remark}[theorem]{Remark}
\numberwithin{equation}{section}
\newcounter{algorithm}

\newcommand{\Q}{\mathbb{Q}}
\newcommand{\Z}{\mathbb{Z}}
\newcommand{\OK}{\mathcal{O}_K}
\newcommand{\Kpos}{K^+_{>0}}
\DeclareMathOperator{\Tr}{Tr}

\title[Unit reducibility in conductor eleven]
{Weak Unit Reducibility of the Eleventh Cyclotomic Field}
\author{Yiyang Chang}
\date{}


\begin{document}

\begin{abstract}
\revision{We prove that the cyclotomic field $K_{11}=\Q(\zeta_{11})$
is unit reducible but not strongly unit reducible.
So every nonzero principal fractional ideal of $K_{11}$ has a
generator that is a shortest nonzero vector.
For $I=h\mathcal O_{K_{11}}$, every shortest nonzero vector
$\alpha\in I$ satisfies $N_{K_{11}/\Q}(\alpha/h)\in\{1,11\}$.
The proof uses Sigrist's Hermite bound for cyclic forms and an
averaging identity for cyclotomic units.
For principal ideals in $K_{11}$, ideal-SVP and the shortest-generator
problem are polynomial-time equivalent.}
\end{abstract}

\maketitle
\pagestyle{plain}

\section{Introduction}\label{1}

The ideal shortest vector problem (ideal-SVP) is to find a nonzero
element of least Euclidean length in an ideal.
For a principal ideal,
the minimum over its generators may be larger than the minimum over
all its nonzero elements. We prove that the two minima agree for every
principal ideal in the eleventh cyclotomic field $K_{11}=\Q(\zeta_{11})$.

\begin{revisionblock}
A number field has the weak shortest-vector/shortest-generator (SVSG)
property if every nonzero principal fractional ideal has a generator
which is a shortest nonzero vector in the canonical Euclidean norm.
It has the strong SVSG property if every shortest nonzero vector of
every such ideal generates that ideal (Definition~\ref{def:svsg}).
\end{revisionblock}

\revision{For CM fields, weak SVSG is equivalent to unit reducibility: every
positive trace form has a unit among its minimal elements
(Proposition~\ref{prop:svsg-equivalence}).}

\revision{Following~\cite{PLSL2025}, \emph{weakly unit reducible} means unit
reducible but not strongly unit reducible; weak SVSG also includes
the strong case.}

\subsection{Related work}\label{1.1}

Unit reducibility was introduced for totally real number fields by
Leibak, Porter and Ling~\cite{LPL2023}. For cyclotomic fields,
Porter--Ling--Sarti--Leibak~\revision{\cite[Theorem~1.4 and
Corollary~1.5]{PLSL2025}} prove strong
unit reducibility for conductors $3,4,5,7,12,15$, and weak unit
reducibility for conductors $8,9$. Their negative results cover
conductors divisible by $16,27,25,49,121$, or by a prime at least $13$.
Our result adds conductor $11$ to the weakly unit reducible cases.
The resulting prime-power classification appears in
Table~\ref{tab:prime-powers}.

Cui--Li--Zhuang~\revision{\cite[Theorem~1]{CLZ2026}} prove that ideals in
\[
 \Z[\sqrt2],\qquad \Z[i],\qquad \Z[\zeta_8],\qquad
 \Z[\zeta_{16}+\zeta_{16}^{7}],
\]
\revision{have generators which are shortest nonzero vectors.} \revision{They apply these results to certain prime ideals in power-of-two
cyclotomic fields. The fourth ring lies in a quartic subfield of $K_{16}$.}

Pan--Xu--Wadleigh--Cheng~\cite{PXWC2021} \revision{use} subfield methods
for prime-ideal SVP, and Porter--Mendelsohn--Ling~\cite{PML2023}
study ideal and module SVP \revision{using decomposition groups.}

\revision{Lyubashevsky, Peikert and Regev~\cite{LPR2013} prove quantum
reductions from worst-case approximate ideal-lattice problems to
average-case Ring-LWE.}

\revision{The algorithm of Cramer--Ducas--Peikert--Regev~\cite{CDPR2016}
takes an arbitrary generator of a principal ideal as input and finds
a short generator by decoding in the logarithmic unit lattice.
Its success bound depends on the distribution of short generators
and the decoding radius. For classical algorithms for class groups
and unit groups, see \cite{BF2014,BPW2026}; for quantum algorithms,
see \cite{EHKS2014,BS2016}. If the field has SVSG, a shortest generator
of a principal ideal is also a shortest nonzero vector.}

\subsection[Main result and proof]{\texorpdfstring{\revision{Main result and proof}}{Main result and proof}}\label{1.2}

\revision{We prove that every positive trace form over $K_{11}$ has a unit
among its minimal elements. If $x$ is a nonunit minimal element,
then $x=u(1-\zeta_{11})$ for a unit $u$ that is also minimal.
Let $I$ be a nonzero principal fractional ideal of $K_{11}$, and let
$\alpha$ be a shortest nonzero vector in $I$.
If $\alpha\mathcal O_{K_{11}}\ne I$, then
\[
 I=\frac{\alpha}{1-\zeta_{11}}\mathcal O_{K_{11}},
 \qquad
 \left\lVert\frac{\alpha}{1-\zeta_{11}}\right\rVert=\lVert\alpha\rVert.
\]
For fixed $K_{11}$, a shortest generator of $I$ can be computed from
$\alpha$ in polynomial time (Proposition~\ref{prop:conversion}).

}\revision{By Sigrist's Hermite bound for $C_{11}$-invariant
forms~\cite[Theorem~B]{Sigrist2000} and AM--GM, the norm of a minimal
element is at most $13$.}
The decomposition of \revision{prime numbers} in $K_{11}$ then leaves only the
norms $1$ and $11$. \revision{For norm $11$,} we prove an averaging
identity for five cyclotomic units. \revision{Using the trace form in
\cite[proof of Theorem~1.4]{PLSL2025}, we exhibit a nonunit minimum
in conductor eleven, which proves that strong unit reducibility fails.}

\revision{With a known generator, the remaining problem is to minimize an
exponential sum over the rank-four logarithmic unit lattice.

The main results are stated in Section~\ref{3} and proved
in Sections~\ref{4}--\ref{5}.
Section~\ref{6} compares CVP with the shortest-generator
problem, states the conditions for bounded-distance decoding (BDD),
and describes a finite search for a shortest generator.
Appendix~\ref{A} verifies the unit identity by polynomial
multiplication.

}

\section{Preliminaries}\label{2}

\revision{For background, see \cite[Chapters~2--6]{Milne2020}.}

\subsection{Number fields, conjugation, and the canonical embedding}\label{2.1}

\revision{Let $F$ be a number field of degree $n$, with ring of integers
$\mathcal O_F$. For $x\in F$, the trace and norm are
\cite[Corollary~2.20]{Milne2020}}
\[
 \Tr_{F/\Q}(x)=\sum_{\sigma:F\hookrightarrow\mathbb C}\sigma(x),
 \qquad
 N_{F/\Q}(x)=\prod_{\sigma:F\hookrightarrow\mathbb C}\sigma(x),
\]
\revision{and $N_F(x)=|N_{F/\Q}(x)|$.
A CM field $F$ is a totally imaginary quadratic extension of a
totally real field $F^+$, with conjugation $x\mapsto\bar x$.
Then $\sigma(\bar x)=\overline{\sigma(x)}$ for every embedding $\sigma$.
Let $F^+_{>0}=\{a\in F^+:\sigma(a)>0\text{ for all }\sigma\}$.
Its elements are called \emph{totally positive}.}

For a CM field of degree $n=2s$, select embeddings
$\tau_1,\ldots,\tau_s$, one from each conjugate pair. The canonical
real embedding is
\[
 \iota_F(x)=\sqrt2\,
 (\operatorname{Re}\tau_1(x),\operatorname{Im}\tau_1(x),\ldots,
  \operatorname{Re}\tau_s(x),\operatorname{Im}\tau_s(x)).
\]
\revision{Then}
\begin{equation}\label{eq:canonical-norm}
 \lVert x\rVert^2:=\lVert\iota_F(x)\rVert_2^2
 =\Tr_{F/\Q}(x\bar x)=2\sum_{j=1}^s|\tau_j(x)|^2.
\end{equation}
\revision{Every nonzero fractional ideal embeds as a full lattice in
$\mathbb R^n$, with finitely many elements in each bounded set.}

\subsection{Ideals, units, and norms}\label{2.2}

\revision{A fractional ideal is a nonzero finitely generated
$\mathcal O_F$-submodule $I\subset F$ with $dI\subseteq\mathcal O_F$
for some integer $d\ne0$. It is principal if $I=h\mathcal O_F$.
Two elements are associated if they differ by a unit in
$\mathcal O_F^\times$; then $h\mathcal O_F=g\mathcal O_F$ if and
only if $g=hu$ for some $u\in\mathcal O_F^\times$.}

For a nonzero integral ideal $\mathfrak a$, set
$N(\mathfrak a)=|\mathcal O_F/\mathfrak a|$, and extend this norm
multiplicatively to fractional ideals. The identities
\begin{equation}\label{eq:prelim-ideal-norm}
 N(x\mathcal O_F)=N_F(x),\qquad
 [I:J]=\frac{N(J)}{N(I)}\quad(0\ne J\subseteq I)
\end{equation}
hold for nonzero $x\in F$, where $[I:J]$ is the additive index.
\revision{For $0\ne x\in\mathcal O_F$, let $M_x$ be the matrix of
$y\mapsto xy$ in an integral basis. Then
\[
 \begin{aligned}
 N(x\mathcal O_F)
 &=[\mathcal O_F:x\mathcal O_F]=|\det M_x|\\
 &=|N_{F/\Q}(x)|=N_F(x).
 \end{aligned}
\]}
Consequently, for $x\in\mathcal O_F$,
\[
 x\in\mathcal O_F^\times\quad\Longleftrightarrow\quad N_F(x)=1.
\]
\revision{The fractional case of \eqref{eq:prelim-ideal-norm} follows by
clearing denominators.}

\revision{Since $\mathcal O_F$ is a Dedekind domain, every nonzero integral
ideal has a unique factorization into prime ideals.}
If $\mathfrak p\cap\Z=\ell\Z$ for a \revision{prime number} $\ell$,
then $\mathcal O_F/\mathfrak p$ is a finite field
with $\ell^f$ elements, where $f$ is the residue degree. Then $N(\mathfrak p)=\ell^f$.

\subsection{Cyclotomic fields}\label{2.3}

For $m\ge3$, let $\zeta_m=\exp(2\pi i/m)$ and $K_m=\Q(\zeta_m)$.
Then $[K_m:\Q]=\varphi(m)$ and
$\mathcal O_{K_m}=\Z[\zeta_m]$. The field is CM, with maximal real
subfield $\Q(\zeta_m+\zeta_m^{-1})$.
For an odd prime $p$,
\[
 \Phi_p(X)=1+X+\cdots+X^{p-1},\qquad
 N_{K_p}(1-\zeta_p)=\Phi_p(1)=p.
\]
\revision{If $\ell\ne p$ is a prime number, the residue degree of each
prime ideal above $\ell$ in $K_p$ is $\operatorname{ord}_p(\ell)$.} Indeed, a primitive $p$th root in characteristic $\ell$ has Frobenius
orbit of length equal to the least positive $f$ for which
$\ell^f\equiv1\pmod p$. The prime $p$ is totally ramified and its
unique prime ideal is $(1-\zeta_p)\mathcal O_{K_p}$.

\revision{Set}
\begin{equation}\label{eq:fixed-field-notation}
 \begin{gathered}
 K=K_{11},\qquad \zeta=\zeta_{11},\qquad \OK=\Z[\zeta],\\
 K^+=\Q(\zeta+\zeta^{-1}),\qquad \varpi=1-\zeta,
 \qquad N(x)=|N_{K/\Q}(x)|.
 \end{gathered}
\end{equation}
The degree is ten, and $1,\zeta,\ldots,\zeta^9$ is an integral basis.
\revision{Let $\sigma_j$ be the embedding with $\sigma_j(\zeta)=\zeta^j$, $1\le j\le10$.}
\revision{The embeddings form five conjugate pairs
$(\sigma_j,\sigma_{11-j})$, $1\le j\le5$.
We use $\sigma_1,\ldots,\sigma_5$, one from each pair.}

\subsection{Units and their logarithmic lattice}\label{2.4}

For a CM field $F$ of degree $2s$, define
\[
 \ell:F^\times\longrightarrow\mathbb R^s,
 \qquad \ell(x)=(\log|\tau_1(x)|,\ldots,\log|\tau_s(x)|).
\]
The logarithmic embedding satisfies
\begin{equation}\label{eq:log-norm-sum}
 \ell(xy)=\ell(x)+\ell(y),\qquad
 \sum_{j=1}^s\ell(x)_j=\frac12\log N_F(x).
\end{equation}
\revision{By Dirichlet's unit theorem~\cite[Theorem~5.1]{Milne2020},
$\ell$ has the roots of unity as its kernel on $\mathcal O_F^\times$,
and its image is a lattice of rank $s-1$:}
\begin{equation}\label{eq:unit-lattice-rank}
 \begin{gathered}
 \Lambda_F=\ell(\mathcal O_F^\times)\subset H,
 \qquad H=\{z\in\mathbb R^s:\textstyle\sum_jz_j=0\},\\
 \operatorname{rank}\Lambda_F=\dim H=s-1.
 \end{gathered}
\end{equation}
\revision{For fundamental units $\epsilon_1,\ldots,\epsilon_{s-1}$,
the vectors $\ell(\epsilon_i)$ form a basis of $\Lambda_F$.
Every unit is $\omega\epsilon_1^{k_1}\cdots\epsilon_{s-1}^{k_{s-1}}$,
with $k_i\in\Z$ and $\omega$ a root of unity.
In particular, $K_{11}$ has unit rank four.}

\section{SVSG and unit reducibility}\label{3}

\revision{All ideals are assumed to be nonzero unless stated otherwise. We use the metric \eqref{eq:canonical-norm}
and the notation \eqref{eq:fixed-field-notation}.}

\subsection{Ideal-SVP, shortest generators, and the SVSG property}\label{3.1}

\revision{For a fractional ideal $I\subset K$, let}
\begin{equation}\label{eq:svsg-lengths}
 \lambda_1(I)=\min_{0\ne y\in I}\lVert y\rVert,
 \qquad
 s_{\mathrm{gen}}(I)=\min_{g\OK=I}\lVert g\rVert
 \quad\text{when }I\text{ is principal}.
\end{equation}
\revision{Clearly, $\lambda_1(I)\le s_{\mathrm{gen}}(I)$.}

The \emph{ideal shortest vector problem} (ideal-SVP) is to find
$\alpha\in I\setminus\{0\}$ with $\lVert\alpha\rVert=\lambda_1(I)$
from a $\Z$-basis of $I$. For a principal ideal, the \emph{shortest
generator problem} is to find $g$ with $g\OK=I$ and
$\lVert g\rVert=s_{\mathrm{gen}}(I)$. \revision{Approximate versions of these problems are discussed in
Section~\ref{6}.}
The \emph{principal ideal problem} (PIP) is to decide whether an ideal
is principal and, if it is, to compute a generator.

\begin{definition}\label{def:svsg}
A principal ideal $I$ has the \emph{(weak) SVSG property} if
$s_{\mathrm{gen}}(I)=\lambda_1(I)$. A field has the SVSG property if
this holds for every nonzero principal fractional ideal in its ring of
integers. A field has the \emph{strong SVSG property} if every shortest nonzero
vector of every principal fractional ideal generates that ideal.
\end{definition}

\revision{Let $I=h\OK$. For ideal-SVP, we minimize $\lVert hx\rVert$
over all nonzero $x\in\OK$. For the shortest generator problem,
we minimize $\lVert hu\rVert$ over all units $u\in\OK^\times$.
We have}
\[
 \lVert hx\rVert^2=\Tr_{K/\Q}(h\bar h\,x\bar x).
\]

\subsection{Trace forms and unit reducibility}\label{3.2}

For $a\in\Kpos$, define the positive trace form
\begin{equation}\label{eq:trace-form}
 q_a(x)=\Tr_{K/\Q}(ax\bar x),
\end{equation}
and
\begin{equation}\label{eq:minima}
 \mu(a)=\min_{0\ne x\in\OK}q_a(x),
 \qquad \mu^*(a)=\min_{u\in\OK^\times}q_a(u).
\end{equation}
\revision{The terms for $\sigma_j$ and $\sigma_{11-j}$ are equal, so}
\begin{equation}\label{eq:embedding-sum}
 q_a(x)=\sum_{j=1}^{10}\sigma_j(a)|\sigma_j(x)|^2
       =2\sum_{j=1}^{5}\sigma_j(a)|\sigma_j(x)|^2.
\end{equation}
\revision{The form $q_a$ is positive definite. The set
$\{x\in\OK:q_a(x)\le q_a(1)\}$ is finite and contains the unit $1$.
Thus both minima in \eqref{eq:minima} exist and are positive.}
For $I=h\OK$, the identity $q_{h\bar h}(x)=\lVert hx\rVert^2$ implies
\begin{equation}\label{eq:ideal-minima}
 \lambda_1(I)^2=\mu(h\bar h),\qquad
 s_{\mathrm{gen}}(I)^2=\mu^*(h\bar h).
\end{equation}

Following~\cite{PLSL2025}, a CM field is called \emph{unit reducible}
if $\mu(a)=\mu^*(a)$ for every totally positive weight, and
\emph{strongly unit reducible} if every minimal element of every such
form is a unit. It is called \emph{weakly unit reducible} if it is
unit reducible but not strongly unit reducible. The same definitions apply
to any CM field $F$, using $\mu_F$, $\mu_F^*$, and $F^+_{>0}$.
The reduction discrepancy of $F$ is defined by
\begin{equation}\label{eq:discrepancy}
 \delta_F=\sup_{a\in F^+_{>0}}\frac{\mu_F^*(a)}{\mu_F(a)}.
\end{equation}
\revision{Then $\delta_F\ge1$, with equality precisely when $F$ is unit reducible.}

\revision{The relation between SVSG and unit reducibility is discussed in
\cite[Section~1.1]{PLSL2025}. We prove their equivalence below.}

\begin{proposition}\label{prop:svsg-equivalence}
Let $F$ be a CM field, equipped with the norm
$\lVert x\rVert^2=\Tr_{F/\Q}(x\bar x)$. Then $F$ is unit reducible
if and only if it has the SVSG property for principal fractional ideals.
\revision{It suffices to consider integral principal ideals.}
\end{proposition}
\begin{proof}
\revision{Suppose that $F$ is unit reducible.} For every principal fractional
ideal $I=h\mathcal O_F$,
\[
 \lambda_1(I)^2=\mu_F(h\bar h)=\mu_F^*(h\bar h)=s_{\mathrm{gen}}(I)^2.
\]
Hence $F$ has the weak SVSG property.

Conversely, assume that $F$ has the weak SVSG property. \revision{Let $2s=[F:\Q]$},
and choose embeddings $\tau_1,\ldots,\tau_s$, one from each conjugate pair.
For $w\in\mathbb R_{>0}^s$, set
\[
 q_w(x)=2\sum_{j=1}^s w_j|\tau_j(x)|^2,\quad
 m(w)=\min_{0\ne x\in\mathcal O_F}q_w(x),\quad
 m^*(w)=\min_{u\in\mathcal O_F^\times}q_w(u).
\]
Fix $w^0\in\mathbb R_{>0}^s$.
In a sufficiently small neighborhood $U$ of $w^0$, there are constants
$0<c<C$ such that $c\le w_j\le C$ for every $w\in U$.
If $q_w(x)\le q_w(1)$, then
\[
 c\lVert x\rVert^2\le q_w(x)\le q_w(1)\le2sC.
\]
\revision{We have $m(w)\le m^*(w)\le q_w(1)$.
The bound above shows that both minima can be computed over fixed finite
subsets of $\mathcal O_F$ for all $w\in U$.
For fixed $x$, $q_w(x)$ is linear in $w$.
Each of $m$ and $m^*$ is the minimum of finitely many continuous functions
on $U$. Both functions are continuous on $U$.}

Suppose that $F$ is not unit reducible. Then some totally positive
$a\in F^+$ satisfies
\[
 m(w^0)<m^*(w^0),\qquad w^0=(\tau_1(a),\ldots,\tau_s(a)).
\]
\revision{By continuity, this strict inequality holds in a neighborhood of $w^0$.} The image of $F$ under
$(\tau_1,\ldots,\tau_s)$ is dense in $\mathbb C^s$: a $\Q$-basis of $F$
identifies this image with an invertible real-linear image of $\Q^{2s}$.
Therefore, we can choose $h\in F^\times$ with each $\tau_j(h)$
sufficiently close to $\sqrt{\tau_j(a)}$ that
\[
 m(w')<m^*(w'),\qquad w'=(|\tau_1(h)|^2,\ldots,|\tau_s(h)|^2).
\]
For $I=h\mathcal O_F$, it follows that
\[
 \lambda_1(I)^2=m(w')<m^*(w')=s_{\mathrm{gen}}(I)^2,
\]
contrary to the weak SVSG property. Thus $F$ is unit reducible.

\revision{Choose an integer $d>0$ with $dI\subseteq\mathcal O_F$.
An element $g$ generates $I$ if and only if $dg$ generates $dI$.
For every $x\in I$, we have $\lVert dx\rVert=d\lVert x\rVert$. So weak and strong SVSG are unchanged when principal
fractional ideals are replaced by principal integral ideals.}
\end{proof}

Strong unit reducibility implies strong SVSG.
\revision{Strong SVSG fails in $K_{11}$
(Remark~\ref{rem:integral-example}).}

\subsection{Gram matrices and Hermite invariants}\label{3.3}

Let $L$ be a lattice of rank $d$ with basis $b_1,\ldots,b_d$ and Gram
matrix $A=(\langle b_i,b_j\rangle)_{i,j}$. A vector with coordinate
column $c\in\Z^d$ has squared length $c^{\mathsf T}Ac$, and
$\det A=\operatorname{covol}(L)^2$.
For a positive definite real symmetric matrix $A$ of size $d$, define
\begin{equation}\label{eq:hermite-definition}
 m(A)=\min_{0\ne c\in\Z^d}c^{\mathsf T}Ac,
 \qquad H(A)=\frac{m(A)}{(\det A)^{1/d}}.
\end{equation}
If $A$ is a Gram matrix of $L$, then $m(A)=\lambda_1(L)^2$.
The Hermite invariant is unchanged by a change of lattice basis or
multiplication of $A$ by a positive scalar.

\revision{A matrix $C\in\mathrm{GL}_d(\Z)$ preserves $A$ if $C^{\mathsf T}AC=A$.
The action of $C_{11}$ generated by multiplication by $\zeta$
preserves each $q_a$. The corresponding Hermite bound is stated in
Section~\ref{4}.}

\subsection{Main results}\label{3.4}

\begin{theorem}\label{thm:main}
Let $K=\Q(\zeta_{11})$ and $\varpi=1-\zeta_{11}$.
For every $a\in\Kpos$,
\[
  \mu(a)=\mu^*(a).
\]
If $x\in\OK\setminus\{0\}$ satisfies $q_a(x)=\mu(a)$,
then
\[
  N(x)\in\{1,11\}.
\]
If the norm is $1$, then $x$ is a unit. If the norm is $11$, then
\[
  \frac{x}{\varpi}\in\OK^\times,
  \qquad
  q_a\!\left(\frac{x}{\varpi}\right)=\mu(a).
\]
\end{theorem}

Strong unit reducibility fails for the following trace form.

\begin{proposition}\label{prop:weakness}
For $a_0=(\varpi\bar\varpi)^{-1}$, one has
$\mu(a_0)=\mu^*(a_0)=10$, and both $1$ and $\varpi$ are minimal
elements. Consequently, $K$ is weakly, but not strongly, unit reducible.
\end{proposition}

For principal ideals, the theorem has the following consequence.

\begin{corollary}\label{cor:shortest-generator}
Every nonzero principal fractional ideal $I$ of $\OK$ admits a generator
$g$ with $\lVert g\rVert=\lambda_1(I)$.
If $\alpha\in I$ is any shortest nonzero vector, then the finite additive
index satisfies
\[
  [I:\alpha\OK]\in\{1,11\},
\]
and the element
\[
  g=
  \begin{cases}
    \alpha, & [I:\alpha\OK]=1,\\[2pt]
    \alpha/\varpi, & [I:\alpha\OK]=11.
  \end{cases}
\]
is a shortest generator.
\end{corollary}

\section{Cyclic trace forms and the norm restriction}\label{4}

\revision{In this section, $a\in\Kpos$.} We prove that every minimal
element of $q_a$ has norm $1$ or $11$.

\subsection{The Gram matrix and its determinant}\label{4.1}

Let $G_a$ be the matrix of
$B_a(x,y)=\Tr_{K/\Q}(ax\bar y)$ in the power basis
$1,\zeta,\ldots,\zeta^9$. Since $a=\bar a$ and trace is invariant under
conjugation, $B_a$ is a real symmetric form. Its matrix entries are
\begin{equation}\label{eq:gram-entries}
 (G_a)_{ij}=\Tr_{K/\Q}(a\zeta^{i-j}),\qquad 0\le i,j\le9.
\end{equation}
If $x=\sum_{j=0}^9c_j\zeta^j$, then $q_a(x)=c^{\mathsf T}G_ac$.

\begin{lemma}\label{lem:determinant}
For every $a\in\Kpos$,
\begin{equation}\label{eq:gram-determinant}
 \det G_a=11^9N_{K/\Q}(a).
\end{equation}
\end{lemma}
\begin{proof}
For every integer $t$,
\[
 \Tr_{K/\Q}(\zeta^t)=
 \begin{cases}10,&11\mid t,\\-1,&11\nmid t.\end{cases}
\]
Then $G_1=11I_{10}-J_{10}$, where $J_{10}$ is the all-ones matrix.
\revision{Its eigenvalues are $1$ and $11$, with multiplicities $1$ and $9$,
so $\det G_1=11^9$.}

Let $M_a$ be the matrix of multiplication by $a$ in the same basis,
using column coordinates.
Since $B_a(x,y)=B_1(ax,y)$, we have
$G_a=M_a^{\mathsf T}G_1$. Over $\mathbb C$, the eigenvalues of $M_a$ are
$\sigma_1(a),\ldots,\sigma_{10}(a)$. \revision{We have}
$\det M_a=\prod_{j=1}^{10}\sigma_j(a)=N_{K/\Q}(a)$.
Taking determinants proves \eqref{eq:gram-determinant}.
\end{proof}

\subsection{The cyclic action and the restricted Hermite bound}\label{4.2}

Since $\zeta$ is a unit, multiplication by $\zeta$ is a $\Z$-linear
automorphism of every fractional ideal of $\OK$. On $\OK$, its matrix
$C$ in the power basis satisfies
\begin{equation}\label{eq:companion-action}
 Ce_j=e_{j+1}\ (0\le j<9),\qquad
 Ce_9=-\sum_{j=0}^{9}e_j,
\end{equation}
where $e_0,\ldots,e_9$ denote the standard coordinate vectors.
\revision{This companion matrix of $\Phi_{11}(X)=1+X+\cdots+X^{10}$
has order $11$ in $\mathrm{GL}_{10}(\Z)$.}

\begin{lemma}\label{lem:cyclic-action}
For every $a\in\Kpos$,
\begin{equation}\label{eq:cyclic-invariance}
 C^{\mathsf T}G_aC=G_a.
\end{equation}
Moreover, if $I=h\OK$ is a nonzero principal fractional ideal, then
multiplication by $\zeta$ acts on its integral basis
$(h,h\zeta,\ldots,h\zeta^9)$ by the same matrix $C$, and preserves the
canonical Euclidean metric on $I$.
\end{lemma}
\begin{proof}
Since $\zeta\bar\zeta=1$, we have
\[
 B_a(\zeta x,\zeta y)
 =\Tr_{K/\Q}(a\zeta\bar\zeta x\bar y)=B_a(x,y).
\]
\revision{Then \eqref{eq:cyclic-invariance} holds. In the basis
$(h,h\zeta,\ldots,h\zeta^9)$, the matrix is again $C$, since
$\zeta(h\zeta^j)=h\zeta^{j+1}$ and $\Phi_{11}(\zeta)=0$.
Taking $a=1$ proves metric invariance on $I$.}
\end{proof}

\begin{remark}\label{rem:galois}
Here $C_{11}$ acts by $x\mapsto\zeta x$, which preserves every ideal
of $\OK$. The Galois automorphisms $\sigma_i:\zeta\mapsto\zeta^i$
\revision{form a group of order $10$ and need not preserve each ideal.}

\end{remark}

\paragraph{Sigrist's Hermite bound.}
For the matrix $C$ in \eqref{eq:companion-action}, define
\[
 H_{C_{11}}=\sup\{H(A):A=A^{\mathsf T}>0,\ C^{\mathsf T}AC=A\}.
\]

\begin{theorem}[Sigrist, Theorem~B]\label{thm:sigrist}
For the standard ten-dimensional representation of $C_{11}$,
\begin{equation}\label{eq:sigrist-constant}
 H_{C_{11}}=\frac4{11^{3/10}}.
\end{equation}
\end{theorem}

This is the $m=11$ case of
\cite[Theorem~B, p.~525]{Sigrist2000}.

\begin{corollary}\label{cor:hermite-bound}
For every $a\in\Kpos$,
\begin{equation}\label{eq:hermite-bound}
 \mu(a)\le4\,11^{3/5}N_{K/\Q}(a)^{1/10}.
\end{equation}
\end{corollary}
\begin{proof}
By Lemma~\ref{lem:cyclic-action}, $G_a$ is $C$-invariant.
Since $m(G_a)=\mu(a)$, Theorem~\ref{thm:sigrist} implies
\[
 \frac{\mu(a)}{(\det G_a)^{1/10}}
 \le\frac4{11^{3/10}}.
\]
Substitution of \eqref{eq:gram-determinant} yields
$\mu(a)\le 4\,11^{9/10-3/10}N_{K/\Q}(a)^{1/10}$, which is
\eqref{eq:hermite-bound}.
\end{proof}

\subsection{A bound on the norm of a minimal element}\label{4.3}

\revision{The norm estimate follows the proof of
\cite[Theorem~1.2]{PLSL2025}, with the Hermite bound for
$C_{11}$-invariant forms.}

\begin{lemma}\label{lem:amgm}
For $a\in\Kpos$ and $0\ne x\in\OK$,
\begin{equation}\label{eq:amgm}
 q_a(x)\ge10\,N_{K/\Q}(a)^{1/10}N(x)^{1/5}.
\end{equation}
\end{lemma}
\begin{proof}
The ten terms $t_j=\sigma_j(a)|\sigma_j(x)|^2$ in
\eqref{eq:embedding-sum} are positive. Their product is
\[
 \prod_{j=1}^{10}t_j
 =N_{K/\Q}(a)\prod_{j=1}^{10}|\sigma_j(x)|^2
 =N_{K/\Q}(a)N(x)^2.
\]
Applying the \revision{AM--GM inequality} proves
\eqref{eq:amgm}.
\end{proof}

\begin{proposition}\label{prop:norm-cutoff}
If $q_a(x)=\mu(a)$, then
\begin{equation}\label{eq:norm-cutoff}
 1\le N(x)\le\frac{42592}{3125}<14.
\end{equation}
In particular, $N(x)$ is an integer at most $13$.
\end{proposition}
\begin{proof}
By Lemma~\ref{lem:amgm} and Corollary~\ref{cor:hermite-bound},
\[
 10N_{K/\Q}(a)^{1/10}N(x)^{1/5}
 \le q_a(x)=\mu(a)
 \le4\,11^{3/5}N_{K/\Q}(a)^{1/10}.
\]
\revision{We divide both sides by $10N_{K/\Q}(a)^{1/10}$ and raise them to the fifth power:}
\[
 N(x)\le\left(\frac25\right)^5 11^3
       =\frac{32\cdot1331}{3125}=\frac{42592}{3125}<14.
\]
Since $N(x)$ is a positive integer, $N(x)\le13$.
\end{proof}

\subsection{Ideals of small norm}\label{4.4}

\begin{lemma}\label{lem:small-ideals}
The only nonzero integral ideals of $\OK$ having norm at most $13$ are
$\OK$ and $\varpi\OK$. Their norms are $1$ and $11$, respectively.
\end{lemma}
\begin{proof}
Let $\mathfrak p$ be a prime ideal of norm at most $13$, lying over the
rational prime $\ell$. Since $N(\mathfrak p)=\ell^f$ for a positive
residue degree $f$, necessarily
$\ell\in\{2,3,5,7,11,13\}$.

\revision{For $\ell\ne11$, the residue degree is $\operatorname{ord}_{11}(\ell)$,
as recalled in Section~\ref{2.3}. The relevant values are}
\[
\begin{array}{c|ccccc}
 \ell&2&3&5&7&13\\ \hline
 \operatorname{ord}_{11}(\ell)&10&5&5&10&10\\
 N(\mathfrak p)&2^{10}&3^5&5^5&7^{10}&13^{10}.
\end{array}
\]
Thus no prime ideal above these five rational primes has norm at most $13$.

For $\ell=11$, we have
$\Phi_{11}(X)=(X-1)^{10}$ in $\mathbb F_{11}[X]$, so there is a unique
prime above $11$, namely $(11,1-\zeta)$. Also
\[
 N(\varpi)=\prod_{j=1}^{10}(1-\zeta^j)=\Phi_{11}(1)=11.
\]
Then $\varpi\OK$ has index $11$ and is the unique prime ideal
above $11$.

Hence the only possible prime factor of a proper ideal of norm at
most $13$ is $\varpi\OK$. Since $N(\varpi^2\OK)=121$, such an ideal
must equal $\varpi\OK$.
\end{proof}

\begin{corollary}\label{cor:minimal-associates}
Every minimal element of $q_a$ is either a unit or of the form
$\varpi\varepsilon$ with $\varepsilon\in\OK^\times$.
\end{corollary}
\begin{proof}
If $x$ is minimal, then $N(x)\le13$ by Proposition~\ref{prop:norm-cutoff}.
By Lemma~\ref{lem:small-ideals},
$x\OK=\OK$ or $x\OK=\varpi\OK$.
In the first case $x$ is a unit; in the second,
$x=\varpi\varepsilon$ for some $\varepsilon\in\OK^\times$.
\end{proof}

\section{Unit reducibility and shortest generators}\label{5}

By Corollary~\ref{cor:minimal-associates}, a nonunit minimal element
has the form $\varpi\varepsilon$ with $\varepsilon\in\OK^\times$.
We prove that $\varepsilon$ is then minimal as well.

\subsection{The cyclotomic-unit averaging identity}\label{5.1}

\begin{lemma}\label{lem:averaging-general}
Let $p$ be an odd prime, $\xi=\exp(2\pi i/p)$, $\pi_p=1-\xi$, and
$r=(p-1)/2$. For $1\le k\le r$, put
$w_k=(1-\xi)/(1-\xi^k)$. Then every $w_k$ is a unit in $\Z[\xi]$, and
\begin{equation}\label{eq:averaging-general}
 \frac1r\sum_{k=1}^{r}w_k\bar w_k
 =\frac{p+1}{12}\pi_p\bar\pi_p.
\end{equation}
The identity holds in $\Q(\xi)^+$.
\end{lemma}
\begin{proof}
\revision{The element $c_k=(1-\xi^k)/(1-\xi)=1+\xi+\cdots+\xi^{k-1}$
is integral. Its numerator and denominator are conjugate, so
$|N_{\Q(\xi)/\Q}(c_k)|=1$. Then $c_k$ and $w_k=c_k^{-1}$ are units.}

Let
\[
 P(T)=1+T+\cdots+T^{p-1}
     =\prod_{k=1}^{p-1}(T-\xi^k).
\]
At $T=1$ we have
\[
 P(1)=p,\qquad P'(1)=\frac{p(p-1)}2,
 \qquad P''(1)=\frac{p(p-1)(p-2)}3.
\]
Evaluating $P'/P$ and its derivative at $T=1$, we obtain
\begin{align}
 A:=\sum_{k=1}^{p-1}\frac1{1-\xi^k}
    &=\frac{p-1}{2},\label{eq:first-reciprocal-sum}\\
 B:=\sum_{k=1}^{p-1}\frac1{(1-\xi^k)^2}
    &=\left(\frac{P'(1)}{P(1)}\right)^2-\frac{P''(1)}{P(1)}
      =\frac{(p-1)(5-p)}{12}.
      \label{eq:second-reciprocal-sum}
\end{align}
For $z\ne0,1$, the identity
\[
 \frac1{(1-z)(1-z^{-1})}
 =\frac1{1-z}-\frac1{(1-z)^2}
\]
implies
\begin{equation}\label{eq:full-reciprocal-sum}
 \sum_{k=1}^{p-1}\frac1{(1-\xi^k)(1-\xi^{-k})}
 =A-B=\frac{p^2-1}{12}.
\end{equation}
\revision{Equation~\eqref{eq:full-reciprocal-sum} also follows from the trace
calculation in \cite[p.~452]{PLSL2025}.}
The terms indexed by $k$ and $p-k$ are equal, so the sum over
$1\le k\le r$ is half the sum in \eqref{eq:full-reciprocal-sum}.
Multiplication by $\pi_p\bar\pi_p/r$ proves
\eqref{eq:averaging-general}. Both sides lie in $\Q(\xi)^+$.
\end{proof}

For $p=11$, the units in the lemma are
\begin{equation}\label{eq:five-units}
 v_k=\frac{\varpi}{1-\zeta^k},\qquad 1\le k\le5,
\end{equation}
and $(p+1)/12=1$. Then
\begin{equation}\label{eq:averaging-eleven}
 \sum_{k=1}^{5}v_k\bar v_k=5\varpi\bar\varpi.
\end{equation}
Appendix~\ref{A} lists polynomial expressions for these
units and verifies \eqref{eq:averaging-eleven} by multiplication.

\subsection{Proof of unit reducibility}\label{5.2}

\begin{proposition}\label{prop:unit-minimum}
Suppose that $a\in\Kpos$ and $x=\varpi\varepsilon$, with
$\varepsilon\in\OK^\times$, satisfies $q_a(x)=\mu(a)$.
Then all five units $\varepsilon v_k$ satisfy
\begin{equation}\label{eq:all-ties}
 q_a(\varepsilon v_k)=\mu(a),\qquad 1\le k\le5.
\end{equation}
In particular, $q_a(\varepsilon)=\mu(a)$.
\end{proposition}
\begin{proof}
Multiply \eqref{eq:averaging-eleven} by
$a\varepsilon\bar\varepsilon$ and take the trace. By linearity,
\begin{align*}
 \sum_{k=1}^{5}q_a(\varepsilon v_k)
 &=\Tr_{K/\Q}\!\left(a\varepsilon\bar\varepsilon
                         \sum_{k=1}^{5}v_k\bar v_k\right)\\
 &=5\Tr_{K/\Q}(a\varepsilon\bar\varepsilon\varpi\bar\varpi)
  =5q_a(x)=5\mu(a).
\end{align*}
\revision{Each summand is at least $\mu(a)$, hence all five equal $\mu(a)$.
Since $v_1=1$, $q_a(\varepsilon)=\mu(a)$.}
\end{proof}

\begin{proof}[Proof of Theorem~\ref{thm:main}]
Let $a\in\Kpos$ and $q_a(x)=\mu(a)$. By
Corollary~\ref{cor:minimal-associates}, either $x\in\OK^\times$ or
$x=\varpi\varepsilon$ with $\varepsilon\in\OK^\times$.
In the second case, Proposition~\ref{prop:unit-minimum} implies
$q_a(x/\varpi)=\mu(a)$. \revision{In both cases, there is a unit $u$ with $q_a(u)=\mu(a)$, so}
$\mu^*(a)=\mu(a)$. The corresponding norms are $1$ and $11$.
\end{proof}

\subsection{A nonunit minimum and the failure of strong reducibility}\label{5.3}

\revision{The weight and residue-class decomposition below occur in
\cite[proof of Theorem~1.4, pp.~451--452]{PLSL2025}.
For conductor eleven, the minimum can be computed explicitly.}

For $a_0=(\varpi\bar\varpi)^{-1}$, we have
$q_{a_0}(x)=\lVert x/\varpi\rVert^2$. To prove
Proposition~\ref{prop:weakness}, we compute the minimum of
$\varpi^{-1}\OK$.

\begin{lemma}\label{lem:coefficient-energy}
For rational numbers $b_0,\ldots,b_{10}$ and
$y=\sum_{j=0}^{10}b_j\zeta^j$,
\begin{equation}\label{eq:coefficient-energy}
 \Tr_{K/\Q}(y\bar y)
 =11\sum_{j=0}^{10}b_j^2-\left(\sum_{j=0}^{10}b_j\right)^2
 =\sum_{0\le i<j\le10}(b_i-b_j)^2.
\end{equation}
\end{lemma}
\begin{proof}
We have $\Tr_{K/\Q}(1)=10$ and $\Tr_{K/\Q}(\zeta^{i-j})=-1$ for distinct $i,j$ in
$\{0,\ldots,10\}$. Expanding $y\bar y$, we find
$\Tr_{K/\Q}(y\bar y)=10\sum_jb_j^2-2\sum_{i<j}b_ib_j$.
\revision{Expanding the squared differences proves
\eqref{eq:coefficient-energy}. Both sides are unchanged by adding a
common constant to the coefficients, since $\sum_{j=0}^{10}\zeta^j=0$.}
\end{proof}

\begin{lemma}\label{lem:integer-minimum}
For every $0\ne y\in\OK$,
\[
 \Tr_{K/\Q}(y\bar y)\ge10.
\]
\end{lemma}
\begin{proof}
\begin{revisionblock}
Let $y=\sum_{j=0}^{10}b_j\zeta^j$, where $b_j\in\Z$ and $b_{10}=0$.
Let $S=\{j:b_j=\min_k b_k\}$ and $r=|S|$.
Since $y\ne0$, the coefficients are not all equal, so $1\le r\le10$.
For $i\in S$ and $j\notin S$, we have $b_j-b_i\ge1$.
By Lemma~\ref{lem:coefficient-energy},
\[
 \Tr_{K/\Q}(y\bar y)
 \ge\sum_{\substack{i\in S\\j\notin S}}(b_i-b_j)^2
 \ge r(11-r)\ge10.\qedhere
\]
\end{revisionblock}
\end{proof}

\revision{The estimate for nonintegral cosets in the following lemma
is the $n=11$ case of Lenstra's closest-point calculation
\cite[Proposition~3.1, p.~461]{Lenstra1975}. We prove this directly.}

\begin{lemma}\label{lem:shifted-minimum}
For every $0\ne y\in\varpi^{-1}\OK$,
\begin{equation}\label{eq:shifted-minimum}
 \Tr_{K/\Q}(y\bar y)\ge10.
\end{equation}
Equality holds for $y=1$ and $y=\varpi^{-1}$.
\end{lemma}
\begin{proof}
If $y\in\OK$, Lemma~\ref{lem:integer-minimum} applies. Otherwise write
$y=x/\varpi$ with $x\in\OK\setminus\varpi\OK$. Since
$\OK/\varpi\OK\cong\mathbb F_{11}$, there exist
$e\in\{1,\ldots,10\}$ and $z\in\OK$ such that
$x=e+\varpi z$. Thus
\begin{equation}\label{eq:shifted-coset}
 y=\frac e\varpi+z.
\end{equation}
The identity
\begin{equation}\label{eq:inverse-uniformizer}
 \frac1{1-\zeta}=\frac1{11}\sum_{j=0}^{9}(10-j)\zeta^j
\end{equation}
follows on multiplication by $1-\zeta$: the numerator becomes
$10-\sum_{j=1}^{10}\zeta^j=11$.
Choose $z=\sum_{j=0}^{9}z_j\zeta^j$ with $z_j\in\Z$, and set
$z_{10}=0$. Equation~\eqref{eq:shifted-coset} becomes
\[
 y=\sum_{j=0}^{10}b_j\zeta^j,
 \qquad b_j=z_j+\frac{e(10-j)}{11}.
\]
\begin{revisionblock}
Each $b_j$ is an integer multiple of $1/11$. For $0\le i<j\le10$,
\[
 b_i-b_j=z_i-z_j+\frac{e(j-i)}{11}\notin\Z,
\]
since $11\nmid e(j-i)$. In particular, $b_i\ne b_j$.
\end{revisionblock}

Arrange the coefficients in increasing order as $d_0<\cdots<d_{10}$.
\revision{Since $d_j-d_i\ge(j-i)/11$ for $i<j$, it follows from
\eqref{eq:coefficient-energy} that}
\begin{align*}
 \Tr_{K/\Q}(y\bar y)
 &\ge\frac1{121}\sum_{0\le i<j\le10}(j-i)^2\\
 &=\frac1{121}\sum_{d=1}^{10}(11-d)d^2
   =\frac{11\cdot385-3025}{121}=10.
\end{align*}
For $y=\varpi^{-1}$, the coefficients in
\eqref{eq:inverse-uniformizer}, with a zero coefficient for
$\zeta^{10}$, are exactly $0,1/11,\ldots,10/11$ in reverse order.
\revision{Equality holds for $y=\varpi^{-1}$.} For $y=1$, it follows
from $\Tr_{K/\Q}(1)=10$.
\end{proof}

\begin{proof}[Proof of Proposition~\ref{prop:weakness}]
All conjugates of $\varpi\bar\varpi$ are positive, so $a_0\in\Kpos$.
For every $0\ne x\in\OK$,
\[
 q_{a_0}(x)=\Tr_{K/\Q}\!\left(
                   \frac{x}{\varpi}\overline{\frac{x}{\varpi}}\right)
 \ge10
\]
by Lemma~\ref{lem:shifted-minimum}. Its equality cases imply
$q_{a_0}(\varpi)=q_{a_0}(1)=10$. Thus $\mu(a_0)=10$.
Since $1$ is a unit, $\mu^*(a_0)=10$ also.
\revision{Since $N(\varpi)=11$, strong unit reducibility fails.
Unit reducibility follows from Theorem~\ref{thm:main}.}
\end{proof}

\begin{revisionblock}
Strong SVSG fails for $J=\varpi^{-1}\OK$: by
Lemma~\ref{lem:shifted-minimum}, both $1/\varpi$ and $1$ are shortest,
whereas $1\OK$ is a proper subideal of $J$.
Multiplying by a positive integer to clear denominators preserves
both shortest vectors and the property of generating the ideal.
\end{revisionblock}

\subsection{Shortest generators of principal ideals}\label{5.4}

\begin{proof}[Proof of Corollary~\ref{cor:shortest-generator}]
Write $I=h\OK$ with $h\ne0$. The map $x\mapsto hx$ is a bijection
from $\OK$ to $I$, and
\[
 \lVert hx\rVert^2
 =\Tr_{K/\Q}(h\bar h x\bar x)=q_{h\bar h}(x).
\]
So, if $\alpha\in I$ is shortest, then $x=\alpha/h$ is minimal for
$q_{h\bar h}$ on $\OK$. Multiplication by $h$ also induces an isomorphism of additive groups
$\OK/x\OK\cong I/\alpha\OK$. Consequently,
\begin{equation}\label{eq:index-norm}
 [I:\alpha\OK]=[\OK:x\OK]=N(x)\in\{1,11\}.
\end{equation}
If the index is $1$, then $\alpha\OK=I$ and $\alpha$ is a shortest
generator. If it is $11$, then $\varepsilon=x/\varpi\in\OK^\times$ and
$q_{h\bar h}(\varepsilon)=q_{h\bar h}(x)$ by
Theorem~\ref{thm:main}. \revision{We have}
\[
 g=h\varepsilon=\frac{\alpha}{\varpi},\qquad
 g\OK=I,\qquad \lVert g\rVert^2=\lVert\alpha\rVert^2.
\]
\end{proof}

\begin{remark}\label{rem:normalization}
For an integral principal ideal $I=h\OK$ and a shortest element
$\alpha\in I$, the normalized element $x=\alpha/h$ satisfies
\[
 N(x)=\frac{N(\alpha)}{N(I)}.
\]
Then $N(\alpha)$ equals $N(I)$ or $11N(I)$, according as
$[I:\alpha\OK]$ is $1$ or $11$. Replacing $h$ by a unit multiple
changes $x$ by a unit multiple and preserves $N(x)$.
\end{remark}

\begin{revisionblock}
\begin{remark}\label{rem:integral-example}
Let $I_0=(11/\varpi)\OK$. Since
\[
 \frac{11}{\varpi}=\prod_{k=2}^{10}(1-\zeta^k)\in\OK,
\]
$I_0$ is an integral principal ideal. By Lemma~\ref{lem:shifted-minimum},
$11/\varpi$ and $11$ are shortest in $I_0=11J$, and
\[
 \left\lVert\frac{11}{\varpi}\right\rVert^2=\lVert11\rVert^2=1210.
\]
The element $11$ does not generate $I_0$, since $[I_0:11\OK]=11$.

\end{remark}
\end{revisionblock}

\subsection{Computing a generator from a shortest vector}\label{5.5}

\begin{revisionblock}
\begin{proposition}\label{prop:conversion}
Let $I\subset K$ be a nonzero principal fractional ideal.
Suppose that a $\Z$-basis of $I$ and an exact shortest nonzero element
$\alpha\in I$ are known. All input coordinates are with respect to the basis $1,\zeta,\ldots,\zeta^9$.
For fixed $K$, a shortest generator of $I$ can be computed in polynomial
time in the binary length of these rational coordinates.
The computation uses rational linear algebra in dimension ten and
at most one division by $\varpi$.
\end{proposition}
\begin{proof}
Let $B$ be the matrix of the input basis in power-basis coordinates.
Let $A_\alpha$ be the matrix of $\alpha,\alpha\zeta,\ldots,\alpha\zeta^9$
in the same coordinates. These elements form a $\Z$-basis of $\alpha\OK$.
Since $\alpha\OK\subseteq I$, we have
\[
 D=B^{-1}A_\alpha\in\mathrm{Mat}_{10}(\Z),\qquad
 m=|\det D|=[I:\alpha\OK].
\]
Corollary~\ref{cor:shortest-generator} implies $m\in\{1,11\}$. Let
\[
 g=\begin{cases}
 \alpha,&m=1,\\
 \alpha/\varpi,&m=11.
 \end{cases}
\]
By the same corollary, $g\OK=I$ and $\lVert g\rVert=\lambda_1(I)$.

The matrices are $10\times10$. If $m=11$, we solve $(1-\zeta)g=\alpha$
for $g$. In the basis $1,\zeta,\ldots,\zeta^9$, this is a system of ten
linear equations over $\Q$. All these computations take polynomial
time in the binary length of the input coordinates.
\end{proof}

For principal ideals in $K$, a shortest generator solves ideal-SVP
(Corollary~\ref{cor:shortest-generator}). To solve the shortest-generator
problem, we can solve ideal-SVP and then apply Proposition~\ref{prop:conversion}.
The two problems are polynomial-time equivalent.

Proposition~\ref{prop:conversion} needs only a basis of $I$ and an exact
shortest nonzero vector $\alpha\in I$. If $\alpha$ is only an approximate
shortest vector, the index $[I:\alpha\OK]$ need not be $1$ or $11$.
\end{revisionblock}

\section{Log-unit lattices and algorithmic applications}\label{6}

\revision{Let $F$ be a CM field of degree $n=2s$ and $I=h\mathcal O_F$.
Suppose that $h$ is known. We want to find a unit
$u\in\mathcal O_F^\times$ such that $\lVert hu\rVert=s_{\mathrm{gen}}(I)$.
We use the logarithmic notation of Section~\ref{2.4}.
}

\subsection{Shortest generators in logarithmic coordinates}\label{6.1}

For $I=h\mathcal O_F$, \revision{let}
\begin{equation}\label{eq:log-target}
 \begin{gathered}
 \nu=N_F(h)=N(I),\qquad
 c=\frac{\log\nu}{n},\\
 t=c\boldsymbol1-\ell(h)\in H,
 \qquad E_s(z)=\sum_{j=1}^s e^{2z_j}.
 \end{gathered}
\end{equation}

\begin{proposition}\label{prop:log-optimization}
With this notation,
\begin{equation}\label{eq:exact-log-objective}
 s_{\mathrm{gen}}(I)^2
   =2\nu^{2/n}\min_{v\in\Lambda_F}E_s(v-t).
\end{equation}
If $F$ has the SVSG property, then
$s_{\mathrm{gen}}(I)^2=\lambda_1(I)^2$.
\revision{For $F=K_{11}$, we compute $\lambda_1(I)^2$ by taking the
minimum over the lattice $\Lambda_F$ of rank four.}
\end{proposition}
\begin{proof}
\revision{Let $v=\ell(u)$. By \eqref{eq:log-norm-sum}, $t\in H$, and}
\[
 \ell(hu)=\ell(h)+v=c\boldsymbol1+(v-t).
\]
\revision{We have}
\[
 \lVert hu\rVert^2
 =2\sum_{j=1}^s|\tau_j(hu)|^2
 =2e^{2c}\sum_{j=1}^s e^{2(v_j-t_j)}
 =2\nu^{2/n}E_s(v-t).
\]
\revision{Every generator of $I$ has the form $hu$.
Taking the minimum over all units $u$ proves \eqref{eq:exact-log-objective}.
Apply Corollary~\ref{cor:shortest-generator} for $F=K_{11}$.}
\end{proof}

If the initial generator is replaced by $h'=hu_0$, with
$\ell(u_0)=v_0\in\Lambda_F$, then
$t'=t-v_0$ and
$\Lambda_F-t'=\Lambda_F-t$.

\subsection{CVP and bounded-distance decoding}\label{6.2}

\revision{Ordinary Euclidean CVP asks for a lattice vector closest to $t$.
In \eqref{eq:exact-log-objective}, we need a vector $v\in\Lambda_F$
for which $E_s(v-t)$ is smallest.}
For example, in the
hyperplane $H\subset\mathbb R^5$, the vectors
\[
 z=(2,-1,-1,0,0),\qquad -z
\]
have the same Euclidean norm, but
$E_5(z)=e^4+2e^{-2}+2$ and
$E_5(-z)=e^{-4}+2e^2+2$ differ. The sublevel sets $\{z\in H:E_s(z)\le B\}$ are convex and need not
be symmetric about the origin. Near zero in $H$,
\begin{equation}\label{eq:local-cvp-expansion}
 E_s(z)=s+2\lVert z\rVert_2^2+O(\lVert z\rVert_2^3),
\end{equation}
because $\sum_j z_j=0$.

\begin{proposition}\label{prop:log-distance-bound}
If a unit $u$ satisfies $\lVert\ell(u)-t\rVert_\infty\le R$, then
\begin{equation}\label{eq:balancing-bound}
 \lVert hu\rVert\le e^R\sqrt n\,\nu^{1/n}
                  \le e^R\lambda_1(I).
\end{equation}
This inequality does not require the SVSG property.
If $F=K_{11}$ and a unit satisfies
\[
 E_5(\ell(u)-t)\le
 \beta\min_{v\in\Lambda_F}E_5(v-t),
 \qquad\beta\ge1,
\]
then $\lVert hu\rVert\le\sqrt\beta\,\lambda_1(I)$.
\end{proposition}
\begin{proof}
Every coordinate of $\ell(u)-t$ is at most $R$, so
$E_s(\ell(u)-t)\le s e^{2R}$. \begin{revisionblock}
By Proposition~\ref{prop:log-optimization},
\[
 \lVert hu\rVert
 =\bigl(2\nu^{2/n}E_s(\ell(u)-t)\bigr)^{1/2}
 \le e^R\sqrt n\,\nu^{1/n}.
\]
\end{revisionblock}

\begin{revisionblock}
For $0\ne y=hx\in I$, we have $N_F(y)=\nu N_F(x)\ge\nu$.
By AM--GM,
\[
 \lVert y\rVert^2
 =2\sum_{j=1}^s|\tau_j(y)|^2
 \ge nN_F(y)^{2/n}
 \ge n\nu^{2/n}.
\]
So $\lambda_1(I)\ge\sqrt n\,\nu^{1/n}$.
\end{revisionblock}

\begin{revisionblock}
For $K_{11}$, \eqref{eq:exact-log-objective} and the hypothesis on $E_5$ imply
\[
 \lVert hu\rVert^2
 =2\nu^{2/n}E_5(\ell(u)-t)
 \le\beta\lambda_1(I)^2.\qedhere
\]
\end{revisionblock}
\end{proof}

\revision{\revision{Let} $r=\lVert\ell(u)-t\rVert_2$. Since
$\lVert\ell(u)-t\rVert_\infty\le r$, Proposition~\ref{prop:log-distance-bound}
implies $\lVert hu\rVert\le e^r\lambda_1(I)$.}
\begin{revisionblock}
If
$r\le\eta\operatorname{dist}(t,\Lambda_F)$, with $\eta\ge1$, then
\[
 \lVert hu\rVert\le
 \exp\!\bigl(\eta\operatorname{dist}(t,\Lambda_F)\bigr)\lambda_1(I).
\]

\end{revisionblock}

\begin{revisionblock}
Let $g$ be a short generator of $I$.
Since $h$ also generates $I$, there is a unit $u_0$ such that $h=gu_0$.
We use logarithmic decoding as in \cite[Section~4]{CDPR2016}.
Let $e=\ell(g)-c\boldsymbol1$. Then
\begin{equation}\label{eq:bdd-target}
 t=-\ell(u_0)-e.
\end{equation}
Let $\rho$ be the guaranteed decoding radius for the norm used by the decoder.
Suppose that $\lVert e\rVert<\rho$. The decoder returns $-\ell(u_0)$.
Let $u\in\mathcal O_F^\times$ satisfy $\ell(u)=-\ell(u_0)$.
Then $u_0u=\omega$ for some root of unity $\omega$, and
\[
 hu=gu_0u=\omega g,\qquad \lVert hu\rVert=\lVert g\rVert.
\]
The output $hu$ is a shortest generator if and only if $g$ is.
If $g$ is random, the probability of an output of the form $\omega g$
is at least $\Pr(\lVert e\rVert<\rho)$.
\end{revisionblock}

\subsection[An algorithm for shortest generators]
{\texorpdfstring{\revision{An algorithm for shortest generators}}{An algorithm for shortest generators}}\label{6.3}

\begin{revisionblock}
The length of a generator from Section~\ref{6.2}
bounds the search for a shortest generator.
\end{revisionblock}

\begin{proposition}\label{prop:finite-unit-search}
Let $\epsilon_1,\ldots,\epsilon_{s-1}$ be fundamental units of $F$
modulo roots of unity. Suppose that these units and a generator $h$
are known. \revision{Let $g_0$ be any known generator of $I$,
and \revision{let}
\[
 B_0=\frac{\lVert g_0\rVert^2}{2\nu^{2/n}},\qquad
 L=\frac12\log B_0.
\]
For $g_0=h$, this becomes $B_0=E_s(-t)$.}
\revision{We can find a shortest generator by checking all
$k\in\Z^{s-1}$ with}
\begin{equation}\label{eq:finite-search-box}
 -(s-1)L\le
 \sum_{i=1}^{s-1}k_i\ell(\epsilon_i)_j-t_j
 \le L\qquad(1\le j\le s).
\end{equation}
\revision{There are finitely many such vectors.
For $F=K_{11}$, a shortest generator is also a shortest vector of $I$.}
\end{proposition}
\begin{proof}
\revision{Let $hu$ be a shortest generator.
Let $v=\ell(u)$ and $z=v-t\in H$.
By Proposition~\ref{prop:log-optimization}, $E_s(z)\le B_0$.
Hence $z_j\le L$ and
$z_j=-\sum_{i\ne j}z_i\ge-(s-1)L$.}
\revision{We have $z_j=\sum_{i=1}^{s-1}k_i\ell(\epsilon_i)_j-t_j$, so
\eqref{eq:finite-search-box} follows. Since the $\ell(\epsilon_i)$ form a
basis of $H$, its inverse image is bounded and contains finitely
many integer vectors. A candidate of least squared length is
a shortest generator. For $K_{11}$, apply
Corollary~\ref{cor:shortest-generator}.}
\end{proof}

\begin{revisionblock}
\paragraph{\emph{Fundamental units for $K_{11}$.}}
Let $v_j$ be as in \eqref{eq:five-units}. We take
\[
 \epsilon_1=v_2,\quad\epsilon_2=v_3,\quad
 \epsilon_3=v_4,\quad\epsilon_4=v_5.
\]
We have
\[
 v_j^{-1}=\frac{1-\zeta^j}{1-\zeta}\qquad(2\le j\le5).
\]
These are the standard cyclotomic units.
Let $h^+(11)$ be the class number of $K^+$. By
\cite[Lemma~2.7 and Theorem~2.8]{CDPR2016},
\[
 \left[\Lambda_K:\sum_{j=2}^5\Z\ell(v_j)\right]=h^+(11).
\]
The field $K^+$ has degree five and discriminant $11^4$.
By Minkowski's bound \cite[Theorem~4.3]{Milne2020}, every ideal
class of $K^+$ contains an integral ideal $\mathfrak a$ with
\[
 N(\mathfrak a)\le\frac{5!}{5^5}\sqrt{11^4}=\frac{2904}{625}<5.
\]
Both $2$ and $3$ have order five in
$(\Z/11\Z)^\times/\{\pm1\}$. They are inert in $K^+$, so there
are no integral ideals of norms $2$, $3$, or $4$.
Thus $\mathfrak a=\mathcal O_{K^+}$ and $h^+(11)=1$.
The four logarithms form a basis of $\Lambda_K$.

If only a proper sublattice of finite index is known, the search
must also cover the other cosets.

\paragraph{\emph{Bounds on the exponents.}}
For $K_{11}$, let $A$ be the first four rows of
$U=[\ell(\epsilon_1)\ \cdots\ \ell(\epsilon_4)]$ and let $t'$ be
the first four coordinates of $t$.
Projection onto the first four coordinates is an isomorphism
$H\to\mathbb R^4$, so $A$ is invertible. \revision{Let} $\kappa=A^{-1}t'$.
For every candidate in \eqref{eq:finite-search-box}, we have
$Ak=t'+z'$ with $\lVert z'\rVert_\infty\le4L$. Then
\[
 \lVert k-\kappa\rVert_\infty
 =\lVert A^{-1}z'\rVert_\infty
 \le4L\lVert A^{-1}\rVert_\infty.
\]
Here $\lVert A^{-1}\rVert_\infty$ is the induced matrix norm.
Choose certified rational bounds
$\underline\kappa_i\le\kappa_i\le\overline\kappa_i$ and
$\overline R\ge4L\lVert A^{-1}\rVert_\infty$ by interval arithmetic.
Let $g_k=h\prod_{i=1}^4\epsilon_i^{k_i}$.
If $g_k$ is a shortest generator, then
\[
 \left\lceil\underline\kappa_i-\overline R\right\rceil
 \le k_i\le
 \left\lfloor\overline\kappa_i+\overline R\right\rfloor
 \qquad(1\le i\le4).
\]
No equality tests at the logarithmic boundaries are needed.

\par\medskip
\noindent\begin{minipage}{\linewidth}
\refstepcounter{algorithm}\label{alg:shortest-generator}
\hrule height .8pt
\vspace{2pt}
\noindent\textbf{Algorithm~\thealgorithm}\enspace A shortest generator in $K_{11}$\par
\vspace{2pt}
\hrule height .4pt
\vspace{2pt}
\small
\noindent\textbf{Input:} Generators $h,g_0$ of $I$ (possibly equal), and the units $\epsilon_1,\ldots,\epsilon_4$ above.\par
\noindent\textbf{Output:} A generator $g$ of $I$ with $\lVert g\rVert=\lambda_1(I)$.\par
\vspace{2pt}
\renewcommand{\arraystretch}{1}
\begin{tabularx}{\linewidth}{@{}r@{\hspace{.5em}}>{\raggedright\arraybackslash}X@{}}
1: & Compute $\nu,t$ from \eqref{eq:log-target}, and $B_0,L$ using $g_0$.\\
2: & Form $A,t'$ and compute the certified bounds $\underline\kappa_i,\overline\kappa_i,\overline R$ above.\\
3: & $g\gets g_0$.\\
4: & \textbf{for each} integer vector $k$ in the box above \textbf{do}\\
5: & \quad $g_k\gets h\prod_{i=1}^4\epsilon_i^{k_i}$.\\
6: & \quad \textbf{if} $\Tr_{K/\Q}(g_k\bar g_k)<\Tr_{K/\Q}(g\bar g)$ \textbf{then}\\
7: & \qquad $g\gets g_k$.\\
8: & \quad \textbf{end if}\\
9: & \textbf{end for}\\
10: & Output $g$.\\
\end{tabularx}
\vspace{2pt}
\hrule height .4pt
\end{minipage}
\par\medskip
If $g_k$ is a shortest generator, then $k$ lies in the box.
The trace comparisons are exact, so the algorithm returns a shortest generator.
We do not claim that this algorithm runs in polynomial time.
\end{revisionblock}

\subsection[Related arithmetic and lattice algorithms]
{\texorpdfstring{\revision{Related arithmetic and lattice algorithms}}{Related arithmetic and lattice algorithms}}\label{6.4}

\begin{revisionblock}
The search in Section~\ref{6.3} requires a generator
of $I$ and fundamental units of $F$.
For $K_{11}$, the units are listed there.

\paragraph{\emph{Computing a generator and units.}}
Biasse and Song \cite[Theorems~1.1 and 1.3]{BS2016} compute compact
$S$-unit generators and solve PIP in quantum polynomial time.
The full proof and bit complexity are in
\cite[Theorem~1 and Corollary~1]{BS2025}.
Their class-group theorem assumes GRH. The two results used here do not.

Biasse and Fieker~\cite[Theorem~6.1]{BF2014} compute class groups
and unit groups in classical subexponential time.
Their analysis assumes GRH and their heuristics on smoothness and
the generation of relations.
De Boer, Pellet-Mary and Wesolowski~\cite[Theorem~1.3]{BPW2026}
start from an LLL-reduced integral basis.
Under ERH, they compute fundamental units in compact representation
in expected subexponential time.
Eisentr\"ager, Hallgren, Kitaev and Song
\cite[Theorem~1.2]{EHKS2014} compute the unit group in quantum
polynomial time. Their output is numerical.
For compact generators, the search in
Section~\ref{6.3} needs certified logarithms
and exact length comparisons. These costs are not included in the bounds above.

\paragraph{\emph{Finding a short generator.}}
A shorter generator reduces the search bound in
Section~\ref{6.3}.
Suppose that $F=\Q(\zeta_m)$ and $m$ is a prime power.
Let $n=\varphi(m)$ and let $C$ be the cyclotomic unit group.
Cramer, Ducas, Peikert and Regev
\cite[Theorem~4.1]{CDPR2016} start from $h=gu$, with $u\in C$.
On success, their algorithm returns $\omega g$ for a root of unity $\omega$.
Their bound on the success probability requires a hypothesis on the
distribution of $g$.
For an arbitrary unit $u$, we use a representative from each coset in
$\mathcal O_F^\times/C$. There are $h^+$ cosets.
For $K_{11}$, $h^+=1$ by Section~\ref{6.3}.
The hypothesis on the distribution of $g$ is still needed.
For arbitrary principal ideals, their quantum algorithm
(Theorem~6.5) returns a generator with SVP approximation factor
$2^{O(\sqrt{m\log m})}=2^{\widetilde O(\sqrt n)}$.
For prime $m$, Ng and Toma~\cite[Theorem~3.1 and Corollary~4.2]{NT2024}
improve the dual-basis bounds under GRH.
Their success estimate assumes a condition on the distribution of the generator.
These asymptotic estimates alone do not imply a success bound for $m=11$.

\paragraph{\emph{General ideals.}}
Cramer, Ducas and Wesolowski~\cite[Theorem~5.1]{CDW2021} solve
approximate ideal-SVP for arbitrary nonzero integral ideals in cyclotomic fields.
Their quantum algorithm assumes GRH and their Assumption~1.
For conductor $m$, the approximation factor is $\exp(\widetilde O(\sqrt m))$.
In their notation, the expected running time is polynomial in $m$,
$h_K^+$ and $\log N(I)$.
With their Assumption~2, $h_K^+\le\operatorname{poly}(m)$, this is
polynomial in the usual input parameters.
For a fractional ideal, we clear the denominators.
Their bit lengths must be counted in the input size.
The output can be used for the search in
Section~\ref{6.3} only if it generates the ideal.
\end{revisionblock}

\subsection{Shortest generators and ideal-SVP}\label{6.5}

\begin{revisionblock}
For a principal ideal in $K_{11}$, solve PIP to find a generator $h$,
then apply Algorithm~\ref{alg:shortest-generator}:
\[
 \text{ideal basis}
 \ \xrightarrow{\ \mathrm{PIP}\ }\ h
 \ \xrightarrow{\ \text{Algorithm~\ref{alg:shortest-generator}}\ }\ hu,
 \qquad \lVert hu\rVert=\lambda_1(I).
\]
The second step solves \eqref{eq:exact-log-objective}.
Proposition~\ref{prop:conversion} computes a shortest generator from
any shortest vector in polynomial time.

Let $F$ be a CM field and let
$D_F=\sup_I s_{\mathrm{gen}}(I)/\lambda_1(I)$.
Here $I$ runs over all nonzero principal fractional ideals of $F$.
We have
\begin{equation}\label{eq:discrepancy-transfer}
 D_F=\sqrt{\delta_F}.
\end{equation}
\begin{revisionblock}
For $I=h\mathcal O_F$,
\[
 \frac{s_{\mathrm{gen}}(I)^2}{\lambda_1(I)^2}
 =\frac{\mu_F^*(h\bar h)}{\mu_F(h\bar h)},
\]
so $D_F^2\le\delta_F$.
The proof of Proposition~\ref{prop:svsg-equivalence} shows that
$m,m^*$ are continuous and $m>0$. Then $m^*/m$ is continuous.
The image of $F$ under $(\tau_1,\ldots,\tau_s)$ is dense in
$\mathbb C^s$, so the vectors $(|\tau_j(h)|^2)_{j=1}^s$,
$h\in F^\times$, are dense in $\mathbb R_{>0}^s$.
Let $a\in F^+_{>0}$. Choose a sequence of these vectors converging to
$(\tau_j(a))_{j=1}^s$. By continuity,
$\mu_F^*(a)/\mu_F(a)\le D_F^2$.
By the definition of $\delta_F$, we have $\delta_F\le D_F^2$.
\end{revisionblock}

If $g$ is a $\rho$-approximation to a shortest generator, then
\[
 \lVert g\rVert\le\rho\,s_{\mathrm{gen}}(I)
 \le\rho\sqrt{\delta_F}\,\lambda_1(I).
\]
For $K_{11}$, $\delta_F=1$, so the approximation factor is $\rho$.
An explicit factor for other conductors requires an upper bound on $\delta_F$.

For the schemes in~\cite{CDPR2016}, a sufficiently short generator
can be used as an equivalent secret key.
To apply this approach, one must compute the principal ideal from the
public data and find a generator within the scheme's length bound.
The success of decoding also depends on the secret distribution.
The result for $K_{11}$ does not imply a general Ring-LWE attack.
\end{revisionblock}

\section[Comparison and further questions]
{\texorpdfstring{\revision{Comparison and further questions}}{Comparison and further questions}}\label{7}

\subsection{Comparison with earlier results}\label{7.1}

Table~\ref{tab:prime-powers} lists the prime-power cases from
Porter--Ling--Sarti--Leibak~\cite[Theorems~1.3--1.4 and
Corollary~1.5]{PLSL2025}, together with the present result for $K_{11}$.
In the notation of \eqref{eq:discrepancy}, they prove
\begin{equation}\label{eq:prior-discrepancy-bounds}
 \delta_{K_{2^k}}\ge 2^{k-3},\qquad
 \delta_{K_{p^k}}\ge\frac{p^{k-1}(p+1)}{12}
 \quad(p\text{ odd prime}),
\end{equation}
and $\delta_{K_m}\ge\delta_{K_d}$ whenever $d\mid m$.
Hence $K_m$ is not unit reducible if $m$ is divisible by $16$, $27$,
$25$, $49$, $121$, or a prime at least $13$. \revision{By \eqref{eq:discrepancy-transfer}, the square roots of these bounds
are lower bounds for $D_F$. These fields fail SVSG.}

\revision{In Table~\ref{tab:prime-powers}, ``weak'' means unit reducible but
not strongly unit reducible. Except for $m=11$, the entries follow from
\cite[Theorem~1.4 and Corollary~1.5]{PLSL2025}.}

\begin{table}[htbp]
\centering
\small
\caption{\revision{Prime-power conductors.}}
\label{tab:prime-powers}
\begin{tabularx}{\textwidth}{@{}lXX@{}}
\toprule
Prime & Unit reducible cases & Cases not unit reducible\\
\midrule
$2$ & $m=4$: strong; $m=8$: weak & $m=2^k$, $k\ge4$\\
$3$ & $m=3$: strong; $m=9$: weak & $m=3^k$, $k\ge3$\\
$5$ & $m=5$: strong & $m=5^k$, $k\ge2$\\
$7$ & $m=7$: strong & $m=7^k$, $k\ge2$\\
$11$ & $m=11$: weak (this paper) & $m=11^k$, $k\ge2$\\
$p\ge13$ & None & $m=p^k$, $k\ge1$\\
\bottomrule
\end{tabularx}
\end{table}

The rational field $K_2=\Q$ is the trivial additional case. For composite
conductors outside this table, $K_{12}$ and $K_{15}$ are strongly unit
reducible by the same theorem. The conjectural list in
\cite[Conjecture~4.1]{PLSL2025} additionally contains
$11,20,21,24$; our argument addresses $11$.
\revision{A field may fail SVSG while particular families of principal ideals
still satisfy $s_{\mathrm{gen}}(I)=\lambda_1(I)$.}

The power-of-two analysis of Cui--Li--Zhuang~\cite{CLZ2026} combines short generators with subfield descent and
treats prime ideals above rational primes $p\equiv3,5\pmod8$ and
$p\equiv7,9\pmod{16}$ in power-of-two cyclotomic fields. \begin{revisionblock}
For $k\ge3$, they prove
\cite[Theorem~2 and Sections~4--5]{CLZ2026} that
\[
 \lambda_1(\mathfrak p)=\sqrt{2^k a_p},\qquad
 a_p^2-2b_p^2=p,
\]
where $(a_p,b_p)$ is the positive solution with least $a_p$ and
$\mathfrak p\subset\Z[\zeta_{2^{k+1}}]$ lies above
$p\equiv7,9\pmod{16}$. The bound $k\ge3$ comes from the $p\equiv7\pmod{16}$ case,
which uses the quartic subfield
$\Q(\zeta_{16}+\zeta_{16}^7)$ of $K_{16}$ and does not cover $k=2$.
\end{revisionblock}

These results use low-degree subfields to treat selected prime
ideals in larger fields. Our theorem applies to all principal ideals
in $K_{11}$.

\subsection{Extensions and further questions}\label{7.2}

\revision{For $K_{11}$, the Hermite bound and prime decomposition leave only
the association classes of $1$ and $1-\zeta_{11}$. The averaging
coefficient in Lemma~\ref{lem:averaging-general} is then one.}

For another odd prime $p$, \revision{let $q_a$ be the corresponding trace form}
on $\Q(\zeta_p)$. The same identity implies, whenever
$x=(1-\zeta_p)\varepsilon$ with $\varepsilon$ a unit,
\[
 \min_{1\le k\le(p-1)/2}q_a(\varepsilon w_k)
 \le\frac{p+1}{12}\,q_a(x).
\]
The corresponding length factor is $\sqrt{(p+1)/12}$.
For other primes, the norm bound may allow further nonunit
association classes. Each of them requires an estimate comparing
its least $q_a$-value with the least value at a unit, valid for every
totally positive $a$. Lemma~\ref{lem:small-ideals} shows that no
further classes occur for $K_{11}$.

\revision{Extension to a larger field would require descent to a principal
ideal of $K_{11}$ with control of shortest lengths, as in the
power-of-two methods of Section~\ref{1.1}. Such a descent requires more than the inclusion $K_{11}\subset L$.}

\appendix
\section{Polynomial verification of the unit identity}\label{A}

The units in \eqref{eq:five-units} have the polynomial expressions
\begin{align*}
 v_1&=1,\\
 v_2&=-(\zeta+\zeta^3+\zeta^5+\zeta^7+\zeta^9),\\
 v_3&=1+\zeta^3+\zeta^6+\zeta^9,\\
 v_4&=1+\zeta^4+\zeta^8,\\
 v_5&=-\zeta-\zeta^6.
\end{align*}
For each $k$, multiplication by $1-\zeta^k$ and reduction modulo
$\Phi_{11}$ yields $1-\zeta$. With $s_j=\zeta^j+\zeta^{-j}$
for $1\le j\le5$, the products are
\[
\begin{array}{c|l}
 k&v_k\bar v_k\\ \hline
 1&1\\
 2&5+4s_2+s_3+3s_4+2s_5\\
 3&4+s_2+3s_3+2s_5\\
 4&3+s_3+2s_4\\
 5&2+s_5.
\end{array}
\]
Since $s_1+\cdots+s_5=-1$, the sum of the right column is
\[
 15+5(s_2+s_3+s_4+s_5)
 =10-5s_1
 =5(1-\zeta)(1-\zeta^{-1}),
\]
which proves \eqref{eq:averaging-eleven}.

\section*{AI acknowledgement}
An LLM was used for computations, editing, and reorganization.

\end{document}